\documentclass[superscriptaddress,prl,twocolumn]{revtex4-1}
\usepackage{graphicx}
\usepackage{amsmath}
\usepackage[dvipsnames]{xcolor}
\usepackage{lineno}

\begin{document}

\title{Non-thermal evolution of dense plasmas driven by intense x-ray fields}

\author{Shenyuan Ren}
\email{shenyuan.ren@physics.ox.ac.uk}
\thanks{Current address: Beijing Jiaotong University, Beijing, China}
\affiliation{Department of Physics, Clarendon Laboratory, University of Oxford, Parks Road, Oxford OX1 3PU, UK}
\author{Yuanfeng Shi}
\affiliation{Department of Physics, Clarendon Laboratory, University of Oxford, Parks Road, Oxford OX1 3PU, UK}
\author{Quincy Y. van den Berg}
\affiliation{Department of Physics, Clarendon Laboratory, University of Oxford, Parks Road, Oxford OX1 3PU, UK}
\author{Muhammad Firmansyah}
\affiliation{Department of Physics, Clarendon Laboratory, University of Oxford, Parks Road, Oxford OX1 3PU, UK}
\author{Hyun-Kyung Chung}
\affiliation{Korea Institute of Fusion Energy, Daejeon, 34133, Republic of Korea}
\author{Elisa V. Fernandez-Tello}
\affiliation{Instituto de Fusi\'{o}n Nuclear, Universidad Polit\'{e}cnica de Madrid, Jos\'{e} Guti\'{e}rrez Abascal 2, 28006 Madrid, Spain}
\author{Pedro Velarde}
\affiliation{Instituto de Fusi\'{o}n Nuclear, Universidad Polit\'{e}cnica de Madrid, Jos\'{e} Guti\'{e}rrez Abascal 2, 28006 Madrid, Spain}
\author{Justin S. Wark}
\affiliation{Department of Physics, Clarendon Laboratory, University of Oxford, Parks Road, Oxford OX1 3PU, UK}

\author{Sam M. Vinko}
\email{sam.vinko@physics.ox.ac.uk}
\affiliation{Department of Physics, Clarendon Laboratory, University of Oxford, Parks Road, Oxford OX1 3PU, UK}
\affiliation{Central Laser Facility, STFC Rutherford Appleton Laboratory, Didcot OX11 0QX, UK}


\begin{abstract}
The advent of x-ray free-electron lasers (XFELs) has enabled a range of new experimental investigations into the properties of matter driven to extreme conditions via intense x-ray-matter interactions. The femtosecond timescales of these interactions lead to the creation of transient high-energy-density plasmas, where both the electrons and the ions may be far from local thermodynamic equilibrium (LTE). Predictive modelling of such systems remains challenging because of the substantially different timescales on which electrons and ions thermalize, and because of the vast number of atomic configurations that are required to describe the resulting highly-ionized plasmas. Here we explore the evolution of systems driven to high energy densities using CCFLY, a non-LTE, Fokker-Planck collisional-radiative code. We use CCFLY to investigate the evolution dynamics of a solid-density plasma driven by an XFEL, and explore the relaxation of the plasma to local thermodynamic equilibrium on femtosecond timescales in terms of the charge state distribution, electron density, and temperature.
\end{abstract}

\pacs{Valid PACS appear here}
\maketitle

\section{Introduction}

Experiments carried out at x-ray free-electron lasers (FELs) over the past decade have shown how intense x-ray fields can produce well-characterized plasmas at high temperatures and densities via isochoric x-ray photoabsortion~\cite{Vinko:2012aa,Fletcher:2015aa,Levy:2015,Ciricosta:2016}. This approach has proven valuable in providing access to physical properties and atomic processes in warm dense matter, of importance to applications across plasma physics~\cite{Ciricosta2012,Vinko2014}, astrophysics and planetary studies~\cite{Eggert:2010aa,Smith:2018aa}, and fusion energy science~\cite{Gaffney:2018,Vinko:2020}. Experimentally, the systems produced are typically studied via x-ray emission or inelastic scattering~\cite{Toleikis:2010} spectroscopy, while the theoretical description is provided by time-dependent atomic kinetics calculations~\cite{doi:10.1063/1.4975712,Ciricosta:2016}.

Because of the short pulse duration of FEL x-ray sources, on the order of 100~fs, the time evolution of the ionization and ionic charge state populations must be modelled directly via collisional-radiative rate equations, and local thermodynamic equilibrium (LTE) cannot be assumed -- such models are thus called non-LTE. In contrast, these same timescales are typically long for the purposes of electron equilibration, and the free electron distribution is often modelled using a Maxwell-Boltzmann distribution function at some evolving (but instantly equilibrating) electron temperature and density. Being able to replace the time evolution of the full electron distribution function by a time-dependent temperature and density is convenient computationally as it leads to a substantial simplification in modelling complexity, but it is also carries significant physical implications. For one, the shape of the electron distribution determines multiple properties that govern the evolution of the atomic kinetics, including photoemission and collisional ionization rates; these can be altered significantly by non-thermal electron distributions.

The importance of properly accounting for electron equilibration dynamics was demonstrated in early ultrafast experiments at the FLASH FEL with XUV pulses a few tens of fs in duration. There, the electron relaxation time was comparable to the typical lifetime of radiative decay, leading to fluorescence emission from a seemingly cold (non-equilibrated) distribution, alongside a bremsstrahlung emission spectrum arising from the non-thermal electrons consistent with a 20-times higher effective temperature~\cite{Vinko:2010,Medvedev:2011}. The contribution from non-thermal electrons to the total emission spectrum is amplified in such experiments due to the combination of low energy XUV photoexcitation ($<100$~eV) and short pulse lengths. In contrast, similar effects are expected to be  smaller at harder x-ray photon energies ($>$1~keV) and for typical SASE pulse durations of 60-80~fs~\cite{Vinko:2012aa,Preston:2017}. Nevertheless, experimental access to both high x-ray intensities and ultra-short pulse lengths ($<$10~fs) will undoubtedly lead to novel x-ray matter interaction regimes where non-thermal electron dynamics dominate the plasma evolution, even on x-ray emission and scattering timescales. This poses a challenge to established modelling approaches, but also provides new opportunities to study ultrafast electron equilibration in a hot-dense plasmas experimentally {\it in situ}.

The effect of non-thermal electrons in extreme conditions can be modelled in various ways, and efforts have been made in both x-ray Thomson scattering~\cite{doi:10.1063/1.5085664}
and x-ray emission spectroscopy investigations via non-thermal atomic kinetics~\cite{DELAVARGA2013542}. A kinetic modelling addition to the particle-in-cell (PIC) simulation suite PICLS~\cite{SENTOKU20086846,PhysRevE.90.051102} allowed for the inclusion of self-consistent radiation transport, photoionization and Auger electron transport, enabling a full kinetic simulation of XFEL-driven plasmas~\cite{PhysRevE.95.063203}. While PIC-based models are very capable in capturing the evolution of the electron distribution in time and in space, they lack the full predictive capability for detailed x-ray emission spectroscopy. In contrast, non-thermal collisional-radiative codes deploying a Vlasov-Boltzmann-Fokker-Planck equation to treat non-Maxwellian electrons~\cite{PhysRevE.100.013202} are able to treat a vast number of ionic charge configurations, essential for predicting x-ray spectroscopy, and allow us to investigate the effect of non-thermal distributions on spectroscopic experiments at FEL facilities.
 
In this context we present CCFLY, a non-thermal, non-LTE collisional radiative code, designed to model the time-dependent evolution of of both electron distributions and ion states interacting with intense x-ray fields on ultra-short timescales. The code follows on from the SCFLY suite~\cite{doi:10.1063/1.4975712,Chung:2007}, itself based on the FLYCHK code~\cite{CHUNG20053}, which has proven successful in modelling and interpreting x-ray FEL experiments over the past decade~\cite{Ciricosta:2016}.

CCFLY allows us to model the plasma evolution by tracking arbitrary electron distribution functions during the atomic-kinetics calculation, foregoing the need to assume a Maxwellian distribution for the free electrons. More precisely, the Fokker-Planck numerical framework is used to describe the evolution of the electron distribution during the elastic electron-electron collisions and account for the electrons released or absorbed during inelastic electron-ion interactions. A self-consistent method is achieved to compute the density and energy balance of the free-electron population together with the ion population dynamics in a partially ionized warm dense plasma. Thus, it becomes possible for us to investigate ionization dynamics in both equilibrated and non-equilibrated systems at high density.
We note here that while CCFLY tracks the time-evolution of the electron distribution and of the ionic charge states, it does not explicitly take into account the motion of ions. As is normally the case in atomic-kinetics modelling, there is no molecular dynamics, and the ions have no associated kinetic energy (and thus no temperature). This is adequate to describe the evolution of x-ray-irradiated samples on femtosecond timescales, but starts to breakdown as the ions are heated. However, ion heating takes place on longer picosecond timescales, normally dictated by electron-phonon coupling dynamics~\cite{Cho:2011,Vinko:2020,Lee:2021,Zhang:2022}.

In this article use CCFLY to study the effect of non-thermal electrons in Al and Mg plasmas irradiated by x-ray pulses with characteristics achievable at current x-ray FEL facilities. We study both normal isochoric heating experimental setups, where the x-ray photon energy is chosen to be higher than the ionization threshold of several ionic charge states, and resonant pumping setups where the x-ray pulse is tuned to a specific bound-bound atomic resonance. In both cases we find that non-thermal electron processes become important for short pulse durations and at high intensities. We quantify the regimes for which instantaneous thermalization assumptions hold, and provide a more rigorous justification for the modelling assumptions employed in previous x-ray spectroscopy experiments at the Linac Coherent Light Source (LCLS)~~\cite{PhysRevE.83.016403,Vinko:2012aa,Ciricosta2012,Cho:2012,Brown:2014tb,Rackstraw:2014,Vinko2015,Ciricosta:2016aa,Ciricosta:2016,Preston:2017,Kasim:2018ty,PhysRevLett.120.055002,Vinko:2020}.

\section{Theory} \label{sec:Theory}

The atomic kinetics modelling approach aims to track the evolution of the charge states in the plasma as it interacts with an intense x-ray pulse. 
The model assumes we have ions in various ionization states, interacting with a classical free electron gas (the ionized free electrons), the density of which is constrained by the requirement for the overall plasma to be charge-neutral. We use a zero dimensional collisional-radiative model, where the transitions between various ion states are driven by electron collisional processes, or are mediated by the photon field. The evolution of the system is described in terms of the changes in the relative population density $N_i(t)$ of each ionic charge state (labeled by the index $i$), and of the free electron distribution $f_E(E,t)$.
A collisionally-induced transition between two ion states given by some cross section $\sigma_{\pm}(E)$ is described in the model by a transition rate per unit volume $S^{\pm}(t)/V$ given by
\begin{equation}\label{eq:ion_trans_rate}
S^{\pm}(t)/V = \int \sqrt{2E/m_e} \; \sigma_{\pm}(E) \; f_E(E,t) \; dE,
\end{equation} 
where $E$ is the kinetic energy of the electron and $m_e$ the electron mass. By computing the transition rates between all the ion states, we can construct a transition matrix $S$.
For each available ion state $i$, its population density $N_i$ can then be determined by the following set of coupled rate equations
\begin{equation}\label{eq:ion_trans_matrix}
\frac{N_i(t+dt)-N_i(t)}{dt} = N_i(t) \left ( \sum_{j \neq i}S_{ji} - \sum_{k \neq i} S_{ik} \right).
\end{equation} 
Because our system is globally neural, the total electron density $N_e$ can be calculated from the ion density and the ionization $z$ as:

\begin{equation}\label{eq:electron_density_from_ion}
N_e(t+dt) = \sum_{i} N_i(t+dt) z_i,
\end{equation} 
where we have summed across all ion states and multiplied them by their ionization $z_i$. From the perspective of energy balance, the ion transitions can either add energy or draw energy from the free electron population. Thus, due to the requirement of energy conservation, the energy change in the free electron distribution is given by:

\begin{equation}\label{eq:energy_release}
\frac{dE_{\rm{kin}}(t)}{dt} = -\sum_{i \neq j} S_{ij}E_{ij} N_{i}(t) + \frac{dE_{rad}(t)}{dt},
\end{equation} 
where $E_{ij}$ is the energy difference in between ionic charge states $i$ and $j$ involved in the transition $i \rightarrow j$. This expression sums up all the energies that are released into, or drawn from, the free electron distribution due to photo-absorption, Auger decay, and collisional interactions. This includes bound-bound radiative transitions, bound-bound collisional transitions, bound-free radiative ionisation, bound-free radiative recombination, bound-free collisional ionisation, bound-free collisional recombination, autoionization, and electron capture.

We will call the model {\it thermal} if we assume that the electron distribution function is Maxwellian, uniquely described by an electron density $N_e$ and a temperature $T$.
For situations where energy is entering the system, for example via x-ray photoabsorption, the temperature and density are instantaneously equilibrated at each time step. Then,
\begin{equation}
\label{eq:electron_temperature}
\frac{3}{2}\, N_e(t) \, k_{\rm B}T(t) = E_{\rm kin}(t).
\end{equation}
In this case, Eqs.~(\ref{eq:electron_density_from_ion}) and (\ref{eq:energy_release}) completely determine the free-electron distribution in terms of the changing temperature and density of the ionized electrons.

In contrast, if a thermal electron distribution cannot be assumed, then the change in the electron distribution needs to be modelled explicitly for all electron energies. This can be accomplished using the Fokker-Plank equation:

\begin{align}\label{eq:electron_temperature_NT}
\frac{\partial f_E(E)}{\partial t}|_{ee} = &-\frac{\partial}{\partial E}[a(E) \; f_E(E)]  \nonumber \\
&+ \frac{1}{2} \frac{\partial ^2}{\partial E^2}[D(E) \; f_E(E)] \nonumber \\
& + I(f_E(E)) + S(E).
\end{align} 


The change in the electron distribution as a function of time is governed by four terms on the right hand side of this equation. The first two terms, containing $a(E)$ and $D(E)$, describe the elastic part of the evolution.
The third term $I(f_E(E))$ denotes all contributions of inelastic collisional processes within the electron distribution, where the free electrons either gain or lose energy to the ionic charge state. The last term $S(E)$ denotes the collection of all source interactions that do not depend on the electron distribution. By capturing the effects on the electron distribution in an inelastic collision operator that considers the photo-ionisation and Auger decay, a self-consistent approach for the ion populations and electrons can be achieved. Then the ionic charge state populations, the electron distribution and its time-derivatives can be fed into an ordinary differential equation (ODE) solver. We refer to this situation as the {\it non-thermal} case. Note that the ion population is never assumed to be thermal, and the time-dependent modelling is purely non-LTE for both the thermal and non-thermal simulation approaches.

\subsection{Atomic data}

CCFLY supports two levels of detail in the atomic data used in the atomic kinetics modelling: superconfigurations and relativistic configurations. In the super-configurational approach, we simply count how many electrons are bound with each quantum number. States are denoted in the $K_aL_bM_cN_d$ notation which means that there are $a$ electrons in the K-shell (n = 1), $b$ electrons in the L-shell (n = 2), $c$ electrons in the M-shell (n = 3) and $d$ electrons in the N-shell (n = 4). The advantage of using superconfigurations is that the dataset can be kept relatively small compared with a more detailed configurational dataset. In contrast, relativistic configurations provide access to more detailed atomic physics modelling, but at a substantially greater computational cost. A relativistic configurations dataset can be generated using the Flexible Atomic Code \cite{doi:10.1139/p07-197}~\cite{doi:10.1063/1.1824864}, via the CFACDB library that provides access to the cFAC-generated databases.

For the study presented in this work we will use the superconfigurational dataset for our atomic kinetics modelling. In contrast, a more detailed atomic model using relativistic configurations calculated by the DHS code~\cite{PhysRevA.21.436} was used for the spectral synthesis where energy levels, statistical weights, and radiative data are applied to
produce frequency-dependent emissivities and opacities. Population distributions of relativistic configurations belonging to a super-configuration (or to a super-shell) were determined by Boltzmann statistics as a function of electron temperature, as described in more detail in ref.~\cite{CHUNG20053,Chung2016,Ciricosta:2016}. 

\subsection{Continuum lowering} 

The effect of continuum lowering, where the binding energies of atomic states are lowered due to high electron and ion density, is accounted for in the atomic kinetic modelling via an ionisation potential depression (IPD) model. The CCFLY code supports two main IPD models: the modified Ecker-Kroll IPD model (EK)~\cite{2013HEDP....9..258P} and the Stewart-Pyatt IPD model~\cite{1966ApJ...144.1203S}. All simulations presented here were carried out with the EK model, to make the results directly comparable to previously published related work.

\subsection{Elastic Electron Collisions}

The elastic part of the evolution of the electron distribution is given by the first two terms in the Fokker Planck equation in Eq.~\ref{eq:electron_temperature_NT}.
The elastic collision terms can alter the shape of the distribution, while preserving particle density and total kinetic energy. Additionally, the distribution function should equilibrate to the correct Maxwellian distribution. In CCFLY, the electron distribution is initialised and evolved in  energy space but in the Fokker-Planck operator, the electrons are evolved in velocity space.

\begin{figure*}
\centering
\includegraphics[width=\textwidth]{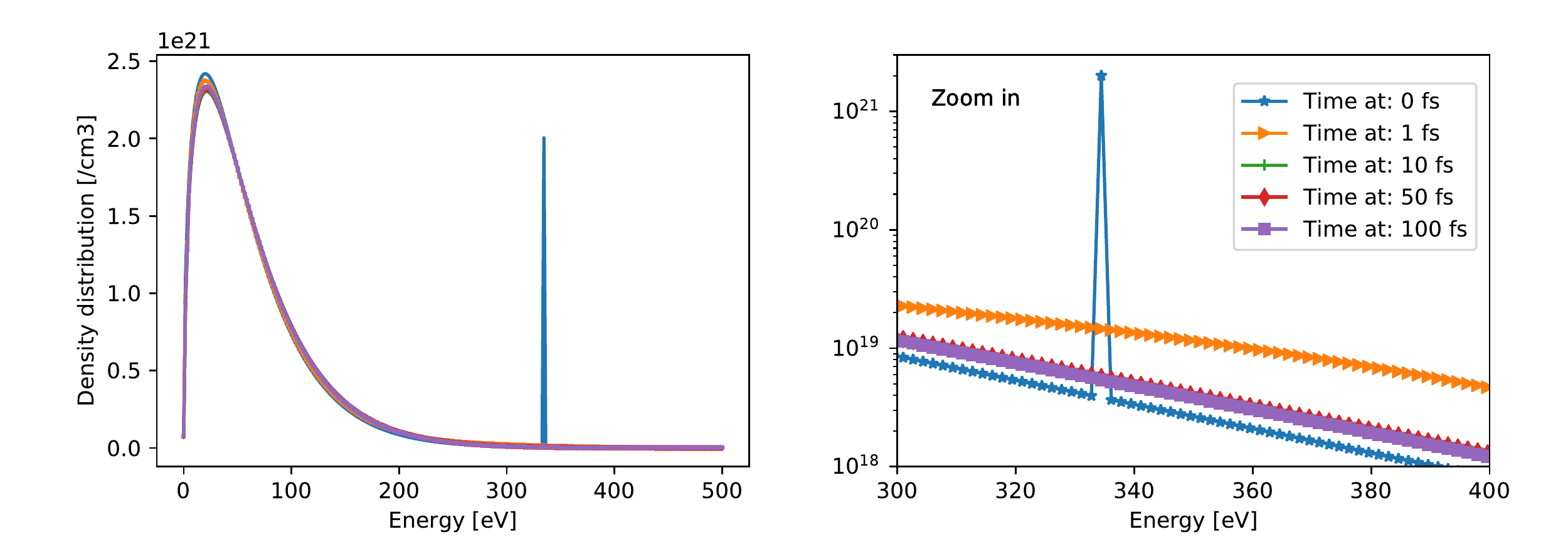}
\caption{Elastic collisional thermalization of the electron distribution function: we initialze a nonthermal distribution consisting of a Maxwellian at a temperature of 40~eV combined with a hot-electron spike at 333~eV. The distribution equilibrates to a hotter Maxwellian distribution on fs timescales.}
\label{fig:ee_collision}
\end{figure*}

First derived by Bobylev and Chuyanov \cite{BOBYLEV1976121}, and cast into a non-relativistic collision operator by Tzoufras et al. \cite{10.1016/j.jcp.2011.04.034}\cite{doi:10.1063/1.4801750}, this formulation is given in integro-differential operator form as:
\begin{equation}
\left(\frac{\partial f_{v}(v)}{\partial t}\right)_{e e}=\frac{4 \pi \Gamma_{e e}}{3} \frac{1}{v^{2}} \frac{\partial}{\partial v}\left[\frac{1}{v} \frac{\partial W\left(v\right)}{\partial v}\right],
\label{eq:pre-factor-a}
\end{equation}
where the function $W(v)$ is given by:

\begin{align}
W\left(v\right)=& f_{v} \int_{0}^{v} f_{u} u^{4} d u \; + \; v^{3} f_{v} \int_{v}^{\infty} f_{u} u d u \; \nonumber \\
& -3 \int_{v}^{\infty} f_{u} u d u \int_{0}^{v} f_{u} u^{2} d u.
\label{eq:pre-factor-b}
\end{align}


The function $W(v)$ contains the time evolution of the electron distribution due to to electron-electron collisions that conserve particle number and kinetic energy. The $\Gamma_{ee}$ is defined as:
\begin{equation}
\Gamma_{e e}=\frac{e^4\ln \Lambda}{4 \pi \varepsilon_0^2 m_{e}^{2}}
\label{eq:gamma_ee}
\end{equation}
with $e$ the electron charge, and $\ln \Lambda$ the Coulomb logarithm. The Coulomb logarithm for electron-electron interactions in hot plasmas is computed with a semi-empirical formula constructed by Huba \cite{Huba2013}:
\begin{equation}
\ln \Lambda=23.5-\ln \left(N_{e}^{1 / 2} T^{-5 / 4}\right)-\left[10^{-5}+(\ln T-2)^{2} / 16\right]^{1 / 2},
\end{equation}
where the density and temperature have units of cm$^{-3}$ and eV, respectively.

For large velocities, Eq.~\eqref{eq:pre-factor-b} can be evaluated through standard discretisation techniques such as the chained trapezoid method. For velocities that are small compared with the characteristic velocity $v_c$ of the distribution, the second and third term in Eq.~\eqref{eq:pre-factor-b} cancel up to $\mathcal{O}(v^5)$. It is necessary to modify the numerical scheme to capture this, in order for the distribution to equilibrate to a Maxwellian. This can be done by writing $W(v_n)$ as a Taylor series $f_{n} \approx f(0)+\left[f_{1}-f(0)\right]\left(v_{n}^{2} / v_{1}^{2}\right)$ and $f(0)=\left(f_{0}-f_{1} v_{0}^{2} / v_{1}^{2}\right) /\left(1-v_{0}^{2} / v_{1}^{2}\right)$. Setting $W\left(v_{n}<v_{c}\right) \equiv W_{s}(v)$, this results in:

\begin{equation}
\begin{aligned} W_{s}(v) = & f_{n} \times\left[f(0) \frac{v_{n}^{5}}{5}+\left(f_{1}-f(0)\right) \frac{v_{n}^{7}}{7 v_{1}^{2}}\right] \\ &+\left[\left(f_{n}-f(0)\right) v_{n}^{3}-\left(f_{1}-f(0)\right) \frac{3 v_{n}^{5}}{5 v_{1}^{2}}\right] \\ & \times \sum_{k=n+1}^{N-1} \frac{1}{2}\left(f_{k} v_{k}+f_{k-1} v_{k-1}\right) \delta_{k-1 / 2}.
\end{aligned}
\label{eq:small_vc}
\end{equation}


Similarly, setting $W\left(v_{n} \geq v_{c}\right) \equiv W_{l}(v)$, we have
\begin{equation}
\begin{aligned} W_{l}(v)=& f_{n} \sum_{k=0}^{n} \frac{1}{2}\left(f_{k} v_{k}^{4}+f_{k-1} v_{k-1}^{4}\right) \delta_{k-1 / 2} \\
&+v_{n}^{3} f_{n} \sum_{k=n+1}^{N-1} \frac{1}{2}\left(f_{k} v_{k}+f_{k-1} v_{k-1}\right) \delta_{k-1 / 2} \\
&-3\left[\sum_{k=n+1}^{N-1} \frac{1}{2}\left(f_{k} v_{k}+f_{k-1} v_{k-1}\right) \delta_{k-1 / 2}\right] \\
& \times\left[\sum_{k=1}^{n} \frac{1}{2}\left(f_{k} v_{k}^{2}+f_{k-1} v_{k-1}^{2}\right) \delta_{k-1 / 2}\right].
\end{aligned}
\label{eq:big_vc}
\end{equation}


We show an example of how this elastic collisional operator relaxes a non-thermal electron distribution to a Maxwellian in Fig.~\ref{fig:ee_collision}. For simplicity, we assume here that only free electrons are involved in the collisional thermalization process, and all interactions with ions (e.g. collisional ionization and recombination) are ignored.
We initialize the electrons to a thermal distribution at 40 eV, and add an additional non-thermal population of electrons with energies around 333 eV. The collisional operator equilibrated the distribution to a new overall temperature of 42 eV, the majority of which takes place on a timescale of a few femtoseconds.

\section{Non-thermal electrons in x-ray isochoric heating}

The capabilities of the CCFLY code allow us to look into the detailed evolution of the electron distribution function in an isochoric heating experiment at an x-ray FEL. In particular, we are interested in exploring what role the non-thermal electrons play in the atomic kinetics of the system, and wish to assess the range of validity of the instantaneous electron thermalization assumption commonly used in the modelling and in the interpretation of experimental results.

\begin{figure*}
\centering
\includegraphics[width=\textwidth]{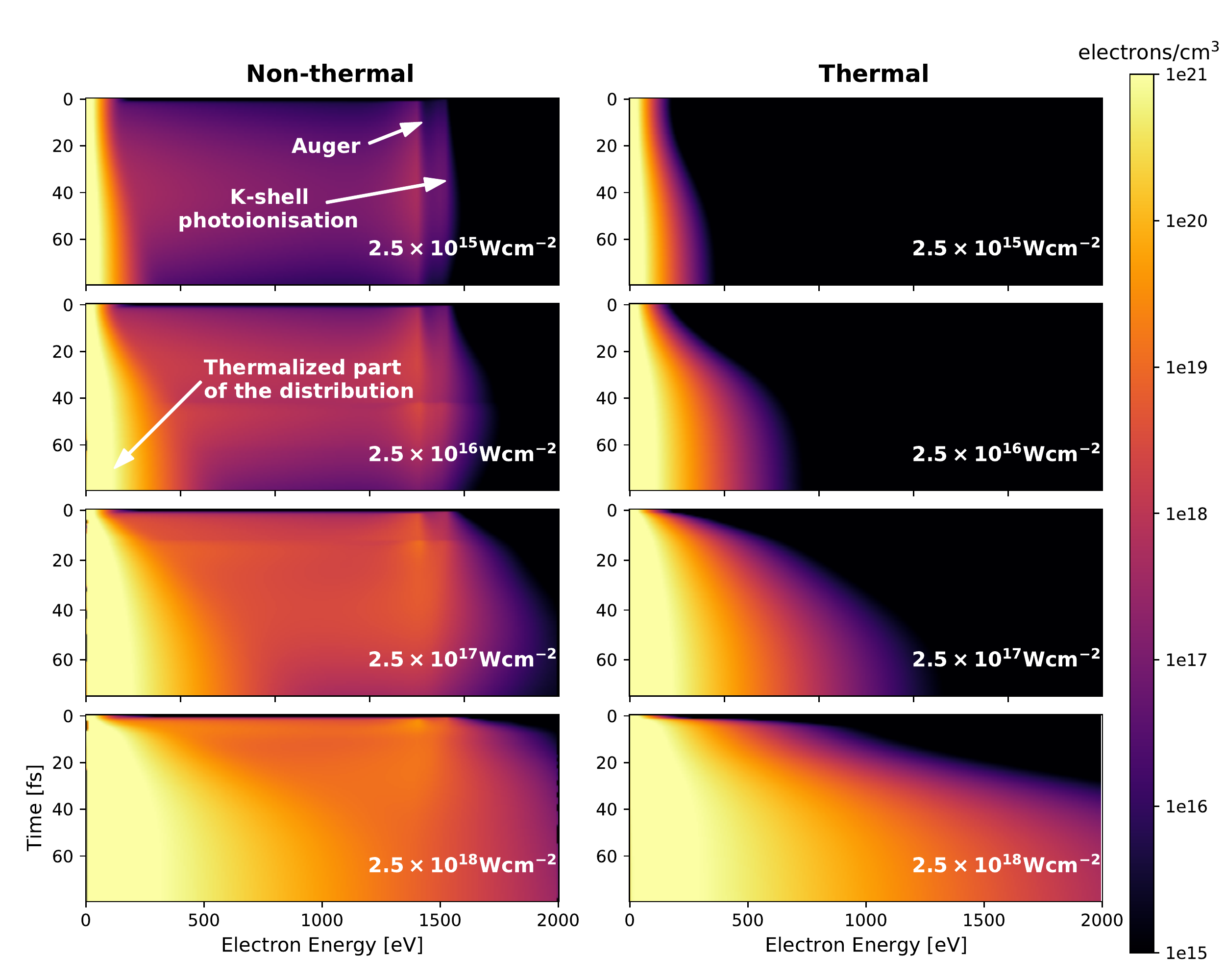}
\caption{Time evolution of the electron distribution in Al irradiated by x-rays at 1590~eV at various peak intensities. The non-thermal electrons generated from photoionization and Auger processes at higher energies are absent in the thermal calculations.
}\label{fig:Al_various_energy_electron_dist}
\end{figure*}

\begin{figure}
\centering
\includegraphics[width=\columnwidth]{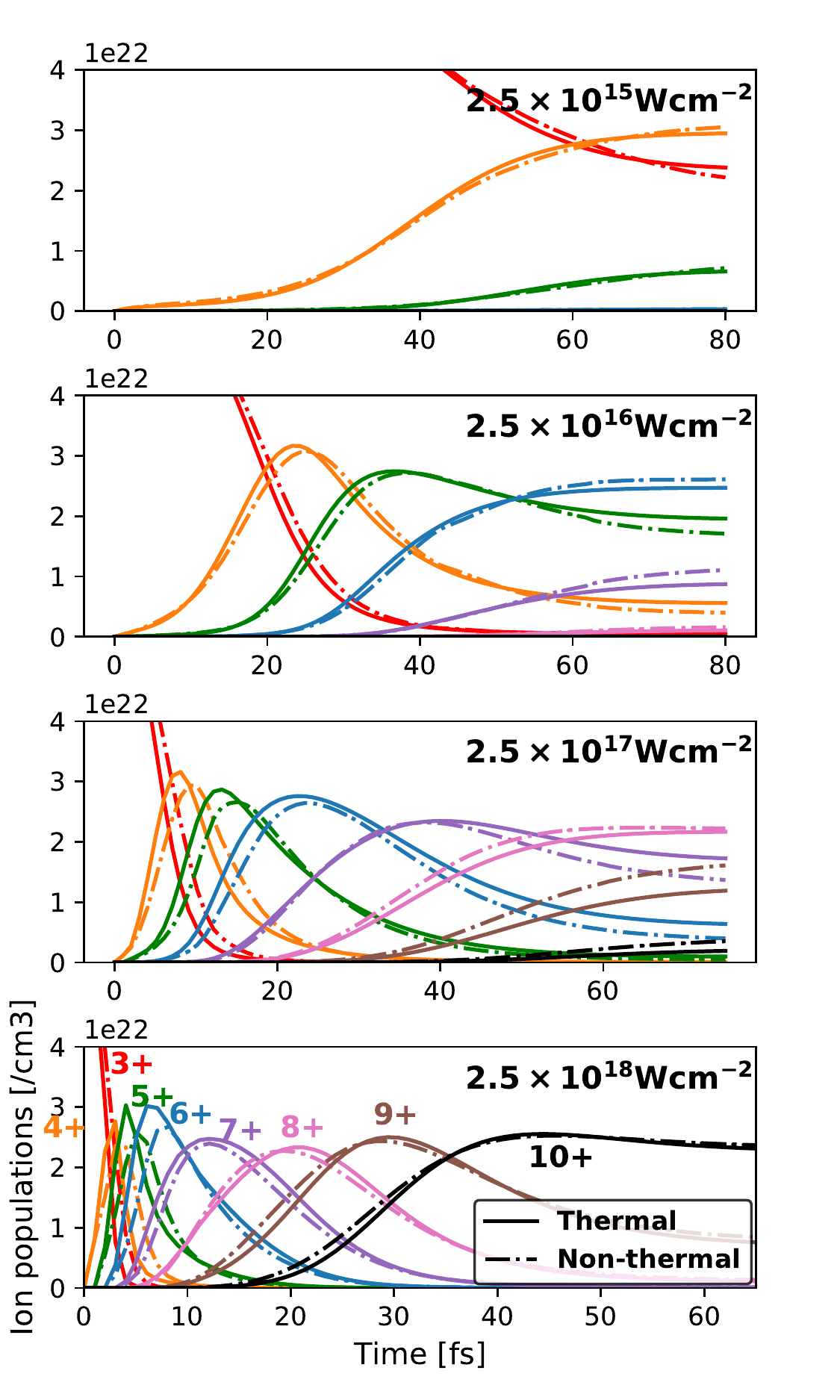}
\caption{Time evolution of the charge state populations in Al irradiated by x-rays at 1590~eV at various peak intensities. Thermal and non-thermal calculations are compared.}\label{fig:Al_various_energy_ion_pop}
\end{figure}

We start by modelling the heating process of a thin Al foil irradiated by x-rays at a photon energy of 1590~eV. This energy was deliberately chosen to be above the Al K-edge (1560~eV) so that the dominant absorption mechanism is K-shell (1s) photoionization.
The x-ray pulse is chosen to be a Gaussian in time with a FWHM duration of 40~fs, and has a spectral bandwidth of 0.4\%, typical of a SASE FEL pulse. Four different intensities are compared, between $2.5\times 10^{15}$~Wcm$^{-2}$ and  $2.5\times 10^{18}$~Wcm$^{-2}$, depositing an ever increasing amount of energy into the target. Experimentally, these intensities would be realized by focusing the x-rays to various spot sizes: focusing the beam to a spot of a few micron in diameter typically yields intensities in region of $10^{17}$~Wcm$^{-2}$ and can be readily achieved on most facilities.

We show the evolution of the electron density in Fig.~\ref{fig:Al_various_energy_electron_dist}, where we compare the results of the thermal and non-thermal models side by side. The x-ray pulse peaks in intensity in the middle of the simulation window, at 40~fs. The thermal simulations simply show a Maxwellian distribution that gets increasingly hotter in time over the duration of the excitation pulse. More deposited energy leads to increasing ionization, and thus the free electron density also increases over time. The temperatures and densities are higher for higher irradiation intensities, since more energy is deposited into smaller volumes, as expected.
The electron distribution function for the non-thermal simulation is visibly different, and can be divided into two components. The first is the bulk part of the free electron distribution, lying at lower energies. This assumes a near-Maxwellian shape because the electron thermalization time is short even on the 100~fs timescales used here (see Fig.~\ref{fig:ee_collision} and related discussion). This bulk component is largely equilibrated, and grows in prominence as the x-rays are absorbed and the sample heats. At higher energies, however, we can see a large energy spread over which non-thermal electrons can be found. These are mostly created by the equilibrating hot electrons generated in the x-ray-matter interaction process, and subsequent atomic kinetics. Two main sources of quasi-mono-energetic hot electrons can be identified: a feature at due to the photoionization of the Al L-shell electrons (at $\approx$1520~eV), and a feature due to the Auger KLL process (at $\approx$1420~eV) that efficiently fills a K-shell core hole. Both features can be readily seen in the figure, but are clearest at the lowest intensities. They are harder to discern at high intensity because their relative populations, compared with the bulk free-electron distribution, decreases with intensity.

Figure~\ref{fig:Al_various_energy_electron_dist} shows the density distribution on a logarithmic scale, and thus amplifies the non-thermal signature. But for the underlying physics, what matters most is the relative contribution of the quasi-thermalized component of the electron distribution in driving the atomic kinetics, compared with the non-thermal hot electron population. To examine this, we plot the populations for ionic charge states $3+$ to $10+$ in Fig.~\ref{fig:Al_various_energy_ion_pop}, in both the thermal and non-thermal case. We immediately note that the results are comparable: the charge state populations do not change dramatically under the instantaneous thermalization approximation, validating this common assumption. Nevertheless, we do see a lag in the time at which certain charge states peak in the population distribution between the thermal and non-thermal calculations. In particular, we note that lower charge states in the more detailed non-thermal calculations tend to appear later in time compared with the thermal prediction, whereas higher charge states appear earlier in time. This is primarily due to the mechanism through which these charge states are created. The lower charge states are more efficiently generated via collisional ionization processes driven by the bulk free electrons; these electrons heat up faster in the thermal calculation as all energy is placed into the thermal distribution instantaneously, whereas in the non-thermal case the bulk temperature rises more slowly as the energy needs to be transferred from the highly energetic photoionization and Auger electrons. In contrast, the higher charge states require substantially higher temperatures to be collisionally ionized, and are thus more efficiently created via collisional interactions with the non-thermal electrons. Even if there are fewer of them available, they have sufficient energy to play an important, and possibly dominant role, in the dynamics of collisional ionization.

In an experiment one cannot normally measure individual plasma charge state populations. However, these populations can be inferred via x-ray emission spectroscopy of K$_{\alpha}$ satellite lines, when the sample is excited by x-rays above the relevant ionization thresholds~\cite{Vinko:2012aa,Vinko:2015}. For charge states photo-pumped below their ionization threshold, the appearance of emission lines can be instead indicative of collisional ionization processes, and thus used to measure collisional rates~\cite{Vinko2015}. For this measurement to be reliable, however, the bulk free electrons must be the dominant contributor to collisional ionization, and the contributions from non-thermal electrons should be small.

\begin{figure}
\centering
\includegraphics[width=\columnwidth]{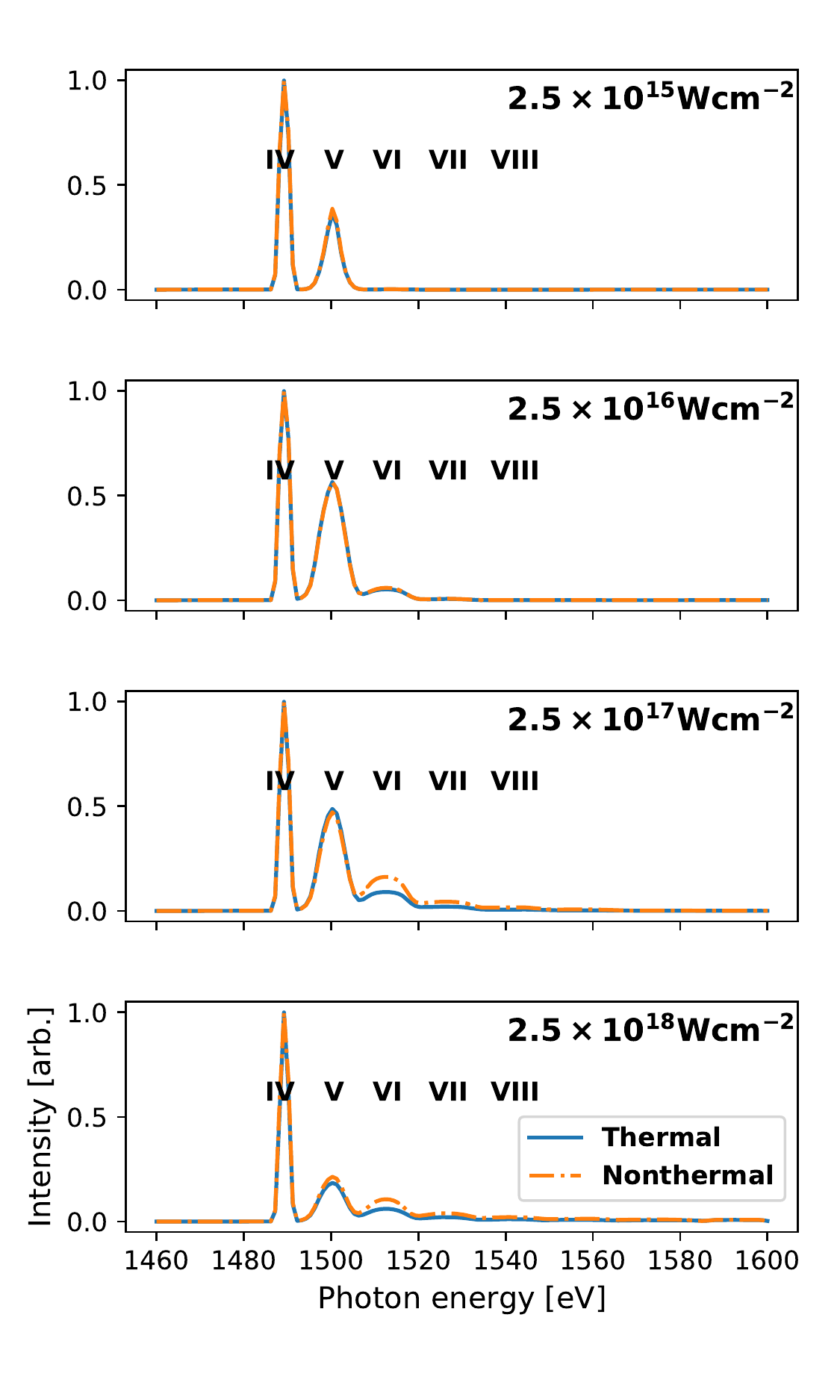}
\caption{Emission spectrum of Al irradiated by x-rays at 1590~eV at various peak intensities, comparing thermal and non-thermal simulations. At this photon energy the emission lines IV and V can be pumped directly via K-shell photoionization, while lines VI to VIII emit because of collisional processes.
}\label{fig:Al_various_energy_spectra}
\end{figure}

We investigate the contribution of non-thermal electrons to the total K$_{\alpha}$ emission spectrum in Fig.~\ref{fig:Al_various_energy_spectra}. The x-ray photon energy is set just above the K-edges of the first two charge states, but below the ionization threshold of higher charge states. Thus only the first two charge states can be revealed directly via K$_{\alpha}$ emission spectroscopy. While higher charge states are still present in the plasma, as illustrated in Fig.~\ref{fig:Al_various_energy_ion_pop}, their $1s$ states cannot be photo-ionized and they therefore cannot relax emitting a K$_{\alpha}$ photon. Despite this, emission from higher charge states can still be observed, because of electron collisional ionization of the L-shell in the presence of a K-shell hole~\cite{PhysRevLett.120.055002}.
Non-thermal electrons affect the emission spectrum in two key ways.
Firstly, because the equilibration of photoionization and Auger electrons is the main heating channel of the bulk free-electron distribution, non-thermal electrons help determine the equilibration timescales and drive the evolution of the charge state population towards LTE. These evolving populations ultimately determine the emission intensity of charge states pumped above their K-edge thresholds. The second mechanism applies to charge states that are pumped below their K-shell ionization threshold. These are primarily driven by collisional ionization of the L-shell electrons, which can be either due to the bulk electrons at some average temperature $k_{\rm B}T\approx 100$~eV, or to the non-thermal electrons at a much higher energy (typically $\approx 10 \, k_{\rm B}T$). Higher electron energies are more efficient at L-shell ionization, and so a system with a large non-thermal electron population and relatively cold bulk distribution will lead to more collisional ionization, and consequently to more intense emission from higher charge states.
We see this play out in the panels of Fig.~\ref{fig:Al_various_energy_spectra}. Because the x-ray pulse length is relatively long compared with the electron relaxation timescale, the thermal and non-thermal modelling predict very similar intensities for the first two emission lines. The charge state population is very close to LTE for intensities up to around 10$^{18}$~Wcm$^{-2}$, and the intensity of the first two lines is simply a reflection of the population of those charge states modulated by the temporal intensity of the x-ray pulse: the second charge state is most prominent at irradiation intensities of around 10$^{16}$~Wcm$^{-2}$, and decreases for lower (insufficient heating) and higher (too much heating) intensities. We note a slight difference in the intensity of the second line at the highest intensities: here the charge state is swept through very quickly, and the instantaneous thermalization assumption starts to breakdown as it underestimates the time needed to establish LTE in the plasma distribution.
The emission intensity of higher lines (VI and above) shows a considerably stronger dependence on the x-ray intensity. For intensities up to several times 10$^{16}$~Wcm$^{-2}$ the thermal and non-thermal predictions match, indicating that the L-shell ionization process described above is indeed dominated by quasi-thermal bulk free electrons. Above this, the number of non-thermal electrons remains sufficiently high to more than double the effective ionization rate, indicating a stark departure from LTE collisional dynamics.

The results shown in Fig.~\ref{fig:Al_various_energy_spectra} are for a single x-ray intensity, and are thus not directly comparable with experimental results that are measured over a broad range of x-ray intensities. However, the average experimental intensity (rather than the often quoted peak intensity) is indicative of the overall irradiation regime, and can be linked to the discussion above. Average intensities typically achieved at the LCLS FEL, such as those fielded for the work on collisional ionization presented in Refs.~\cite{Vinko2015,PhysRevLett.120.055002}, is on the order of~$10^{16}$~Wcm$^{-2}$.

\subsection{The effect of x-ray pulse length}

\begin{figure}
\centering
\includegraphics[width=\columnwidth]{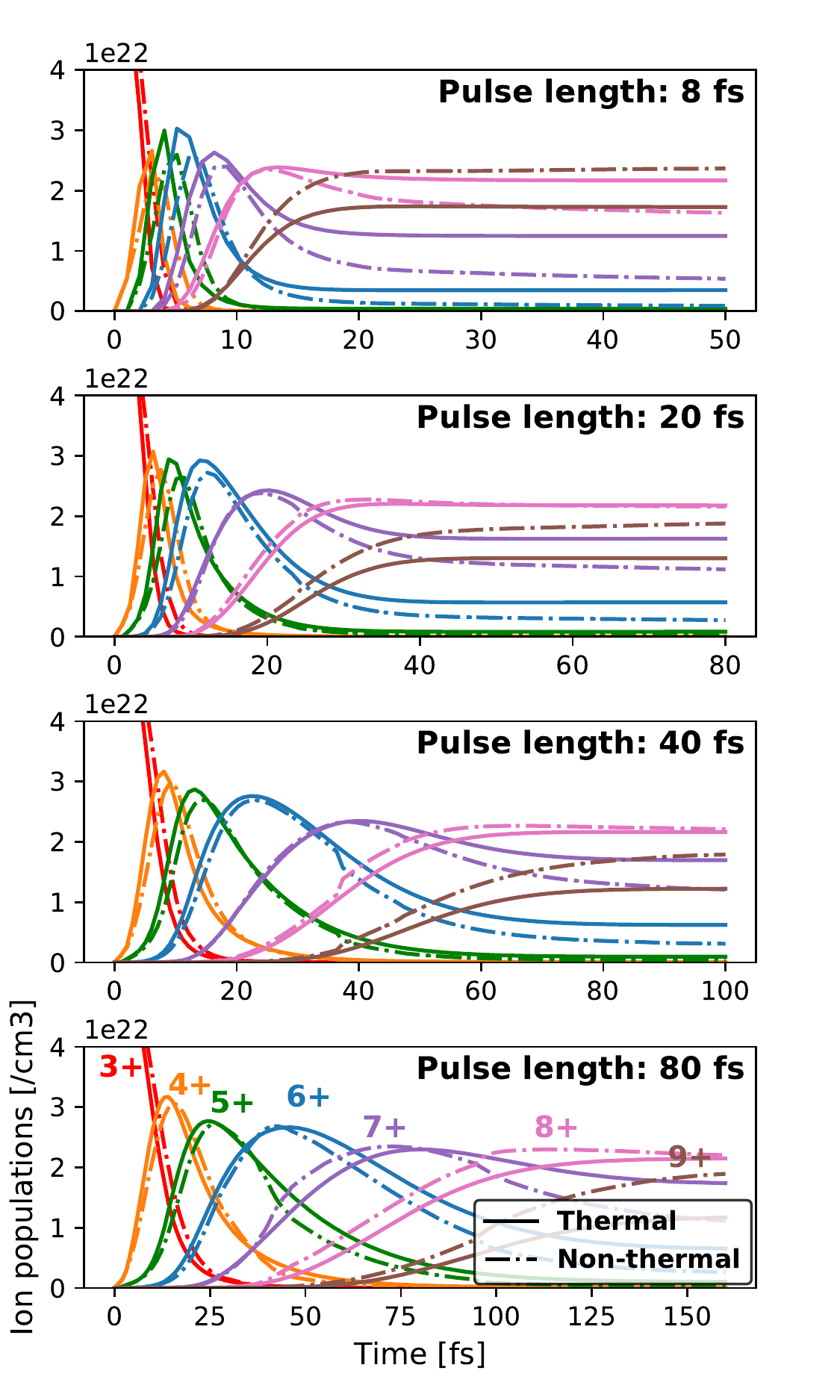}
\caption{Evolution of the charge state populations in Al irradiated at constant energy for various pulse lengths. Thermal and non-thermal calculations are compared.} \label{fig:Al_same_energy_ion_pop}
\end{figure}

\begin{figure}
\centering
\includegraphics[width=\columnwidth]{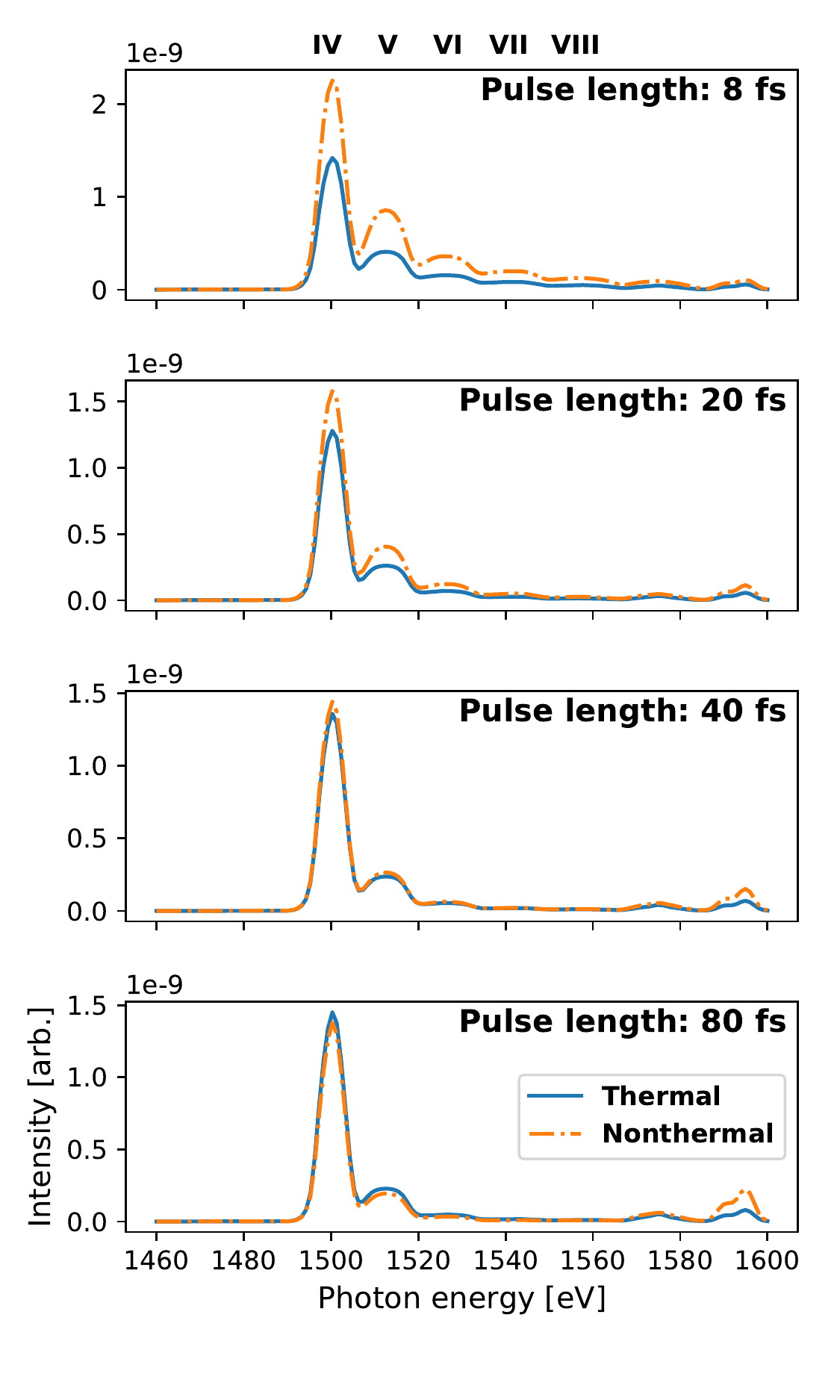}
\caption{Emission spectra of Al irradiated at constant energy for various pulse lengths. Differences in the emission intensity can be seen for pulse durations around 20~fs and below.}\label{fig:Al_same_energy_spectra}
\end{figure}

We have seen that for a given x-ray pulse duration, the effect of focusing the x-ray beam to achieve increasing intensities is to increase the relative contribution of non-thermal electrons to the evolution of the plasma, and thus to the observed emission spectrum. We now turn our attention to the effect of pulse length. Changing the pulse length is another way to change the intensity on target, but it adds an additional layer of complexity because the pulse duration also gates the observation timescale of the system. This is primarily because we only observe x-ray emission in the K$_{\alpha}$ window when $1s$ electrons can be ionized, and the overall temperatures reached in a typical microfocusing experiment ($\approx 100$~eV) are insufficient to drive collisional K-shell ionization to any notable extent. Thus, emission is only observed during the x-ray pulse, and for a short time after depending on the femtosecond core-hole lifetime. By changing the duration of the x-ray pulse, while maintaining a constant pulse energy, we thus not only modify the x-ray intensity seen by the plasma, but also change the period over which we view the evolution of the charge state distribution. We show the simulated results for a Al plasma in Figs.~\ref{fig:Al_same_energy_ion_pop} and \ref{fig:Al_same_energy_spectra}. 
For these simulations we have assumed a pulse energy of 1~mJ, and a spotsize on-target of 12~$\mu$m$^2$, so that a pulse length of 80~fs corresponds to an x-ray intensity of about 10$^{17}$~Wcm$^{-2}$. These are typical values achievable at FEL facilities today~\cite{Vinko:2012aa}. The pulse energy is then kept constant, and the pulse duration lowered down to 8~fs.

For the longer pulse lengths of 40 and 80~fs we see little difference in the computed emission spectrum for the thermal and non-thermal cases, and a lag in the creation of various charge states analogous with that described in the previous section. In these cases, the observation window is sufficiently long for the evolution of the change state populations to agree well with the assumption of instantaneous electron thermalization. The overall population of the L-shell states is collisionally driven by the equilibrated bulk free electron distribution, and the charge states evolve as if in near-LTE for each point in time. For pulse durations of 20~fs and less, these assumptions start to breakdown, and the calculated spectrum assuming instantaneous electron thermalization underestimates the emission intensity.

\begin{figure}
\centering
\includegraphics[width=\columnwidth]{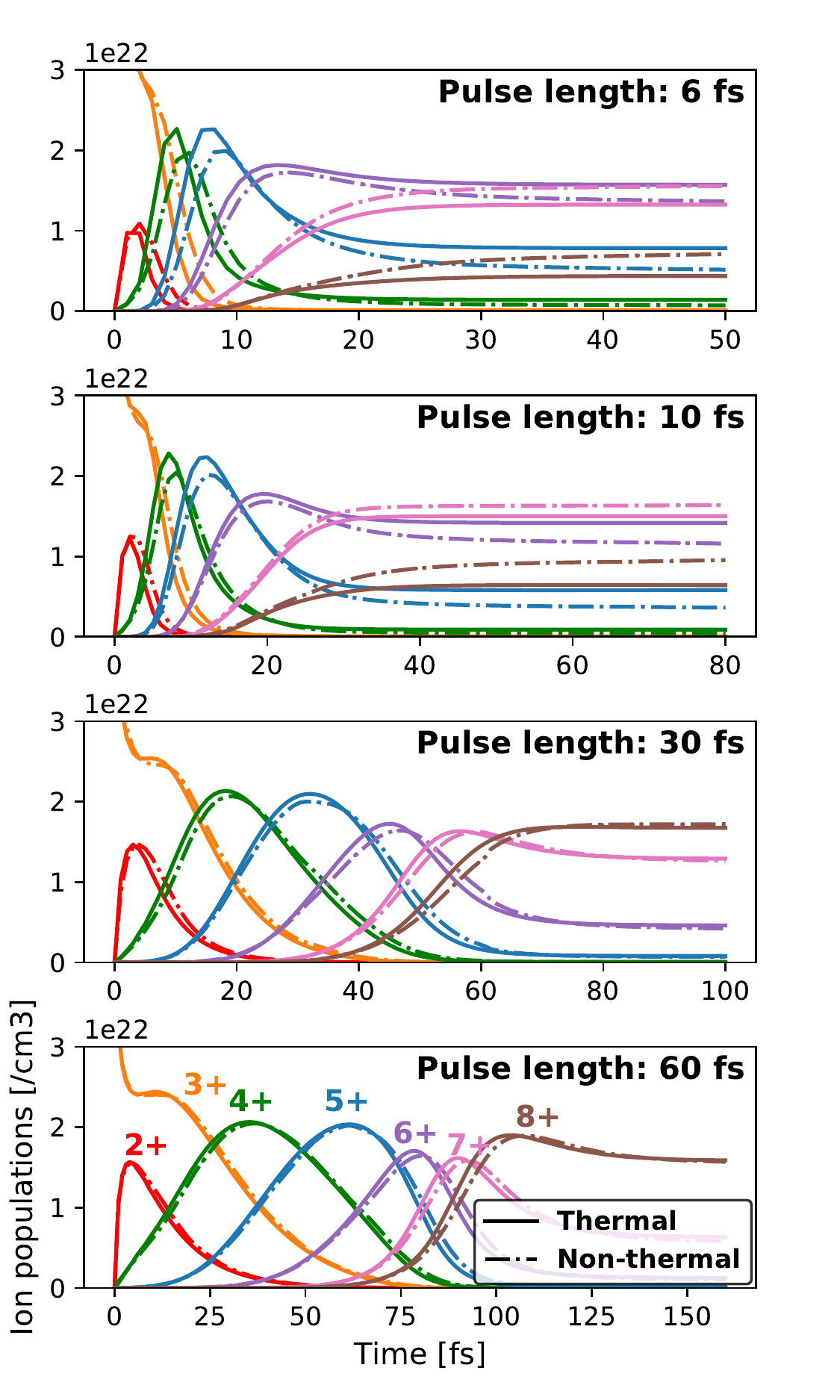}
\caption{Evolution of the charge state populations in Mg irradiated on-resonance at 1304 ~eV, at constant energy for various pulse lengths. Thermal and non-thermal calculations are compared.}\label{fig:Mg_same_energy_ion_pop}
\end{figure}

\subsection{Non-thermal electrons in resonant spectroscopy}

\begin{figure}
\centering
\includegraphics[width=\columnwidth]{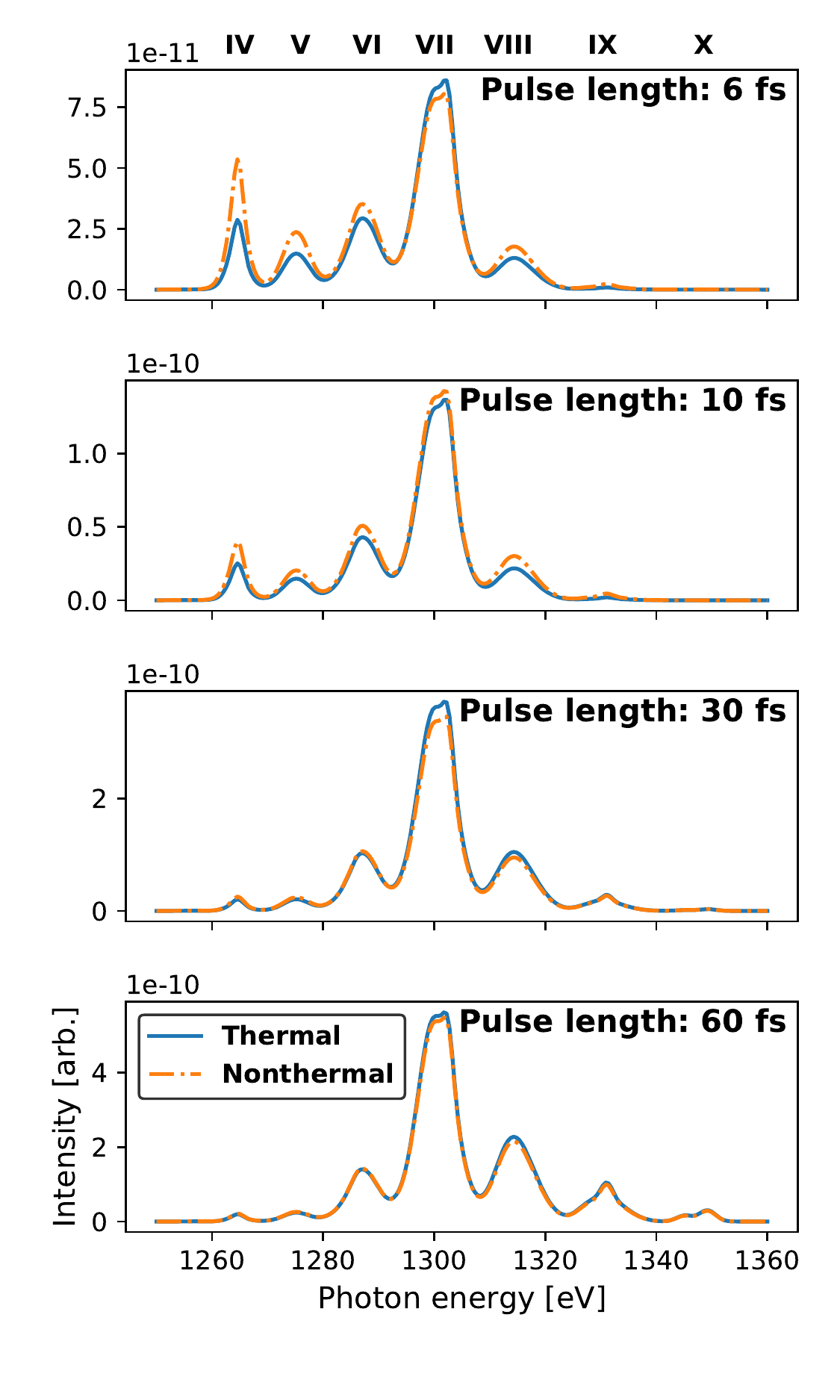}
\caption{Emission spectra of Mg irradiated on-resonance at 1304 ~eV, at constant energy for various pulse lengths. The emission lines on two sides of the resonant VII K$\alpha$ line are due to collisional ionization (right of the K$\alpha$) and three-body recombination (left of the K$\alpha$). Non-thermal electrons start to affect the spectrum for pulse lengths below 30~fs.}\label{fig:Mg_same_energy_spectra}
\end{figure}

Recently, van den Berg {\it et al.} proposed driving a resonant $1s\rightarrow2p$ transition in Mg to measure collisional ionization rates in dense plasmas~\cite{PhysRevLett.120.055002}. The principle is straightforward: an x-ray pulse form an FEL is tuned to resonantly drive a specific inner-shell K$_{\alpha}$ transition, and a spectroscopic measurement is performed to look for emission from neighbouring K$_{\alpha}$ emission lines. Because of the narrow bandwidth of the FEL beam, the x-ray pulse can only excite a single K$_{\alpha}$ line, and so the observation of multiple lines is indicative of L-shell collisional processes taking place {\it during} the lifetime of a K-shell core hole. This method thus allows for a measurement of an ultra-fast collisional ionization rate. However, the real objective is not so much to infer a rate, but rather to measure a collisional cross section. This can be accomplished using the expression in Eq.~(\ref{eq:ion_trans_rate}), under the caveat that the collisions are predominantly due to an interaction of the ion with the bulk free-electron distribution at some known temperature and density. But if the electron distribution is non-thermal, then it is easy to see that the cross section cannot uniquely be extracted from the integral expression of Eq.~(\ref{eq:ion_trans_rate}). Estimating the effect of non-thermal electrons is then important to guide the reliable interpretation of such experiments, especially since we have shown that the contribution of non-thermal electrons can be significant with increasing x-ray intensity or decreasing x-ray pulse length.  

For this investigation we set out simulation parameters to match the conditions fo Ref.~\cite{PhysRevLett.120.055002}: an optically thin Mg plasma is pumped at a photon energy of 1304~eV, corresponding to the resonance energy of the $1s \rightarrow 2p$ transition energy in a 7+ Mg ion. At the nominal pulse duration of 60~fs the intensity on target is $10^{17}$~Wcm$^{-2}$. We show the evolution of the ionic populations as a function of time, using the thermal and a non-thermal models, in Fig.~\ref{fig:Mg_same_energy_ion_pop}. The related emission spectra are shown in Fig.~\ref{fig:Mg_same_energy_spectra}. Here we see that the assumption of instantaneous thermalization holds well for nominal pulse durations, above $\sim$30~fs, providing a more rigorous justification of the assumptions made in the work on collisional ionization rates. For shorter pulse durations, however, the contribution of non-thermal electrons starts to become significant, and even dominant for pulse lengths below 10~fs. At these short pulse lengths, modelling the full evolution of the electron distribution function becomes essential to produce atomic kinetics simulations with predictive capability. 

\section{Conclusions}

We have presented the atomic kinetics simulation package CCFLY, which can model the non-thermal (electron) and non-LTE (ion) evolution of matter interacting with intense x-ray radiation on ultra-short timescales. We have used these capabilities to investigate the evolution of solid-density systems, heated isochorically to high temperatures via femtosecond FEL pulses, and assess the role of non-thermal electron distributions. We find that in mid-Z systems such as Al and Mg, non-thermal electrons become important in determining the short-time plasma kinetics and electron dynamics for intensities above $10^{17}$~Wcm$^{-2}$, and for pulse lengths below 10~fs. 

To date, the use of Fokker-Planck models within collisional-radiative simulations to interpret experiments at free-electron lasers has been limited, largely due to the substantial computational cost the method requires. Whilst we have shown that this is often justifiable for experiments using a typical SASE FEL pulse on the timescales of order 100~fs, ongoing developments enabling the routine generation of energetic few-fs and sub-fs pulses, and of ultra-high intensities via narrow bandwidth self-seeding and x-ray nanofocusing, are increasingly accessing x-ray regimes where the direct modelling of non-thermal electron relaxation will be essential to retain predictive power in the modelling of extreme x-ray matter interactions.

\section{Acknowledgments}
S.R., M.F.K., J.S.W. and S.M.V. acknowledge support from the UK EPSRC grant EP/P015794/1 and the Royal Society.
S.M.V. is a Royal Society University Research Fellow.

\bibliographystyle{apsrev4-1}
\bibliography{ref}

\end{document}